\documentclass[sn-mathphys-num]{sn-jnl}
\usepackage{graphicx}%
\usepackage{multirow}%
\usepackage{amsmath,amssymb,amsfonts}%
\usepackage{amsthm}%
\usepackage{mathrsfs}%
\usepackage[title]{appendix}%
\usepackage{xcolor}%
\usepackage{textcomp}%
\usepackage{manyfoot}%
\usepackage{booktabs}%
\usepackage{algorithm}%
\usepackage{algorithmicx}%
\usepackage{algpseudocode}%
\usepackage{listings}%
\usepackage[numbers]{natbib}
\usepackage{subcaption}

\theoremstyle{thmstyleone}%
\theoremstyle{thmstyletwo}%

\theoremstyle{thmstylethree}%

\begin{document}

\title{Impact of spatial curvature on quantum Otto engines}


\author[1]{\fnm{Somayeh} \sur{Kourkinejat}}\email{s.koorkinejat@gmail.com}

\author*[1,2]{\fnm{Ali} \sur{Mahdifar}}\email{a.mahdifar@sci.ui.as.ir}

\author*[3]{\fnm{Ehsan} \sur{Amooghorban}}\email{ehsan.amooghorban@sku.ac.ir}

\affil[1]{\orgdiv{Physics Department}, \orgname{University of Isfahan}, \orgaddress{\street{ Hezar Jerib St. Isfahan}, \city{ Isfahan}, \postcode{81764-73441},  \country{ Iran}}}

\affil*[2]{\orgdiv{Quantum Optics Group}, \orgdiv{Department of Physics}, \orgname{University of Isfahan}, \orgaddress{\street{ Hezar Jerib St. Isfahan}, \city{ Isfahan}, \postcode{81764-73441}, \country{ Iran}}}

\affil*[3]{\orgdiv{Physics Department}, \orgdiv{ Faculty of Science}, \orgname{Shahrekord University}, \city{ Shahrekord}, \postcode{ 8818634141}, \country{ Iran}}


\abstract{In this paper, we consider a quantum Otto cycle with a  quantum  harmonic oscillator on a circle as its working substance. Since the eigen-energies of this oscillator depend on the curvature of the circle, this model, as an analog model, enables us to investigate the curvature effects of the physical space on properties of quantum heat  engines. We assume that two classical hot and cold thermal baths are located at places with different curvatures. We calculate the curvature-dependent work and heat in our Otto cycle with a particular emphasis on how curvature affects  it's thermal efficiency. By adjusting the curvature difference between the locations of the thermal baths, we demonstrate that the efficiency of our heat engine can reach the Carnot limit.}


\maketitle

\section{Introduction}\label{sec1}

Einstein’s theory of general relativity indicates that massive objects  change the geometric nature, and thus, the  curvature of the surrounding space-time,  leading to some curvature-dependent effects \cite{R29,R54}. Furthermore, the equivalence principle provides a clear insight into the impact of classical gravitational fields on specific physical experiments \cite{Mahdifar:13}.
Although these curvature effects predicted by general relativity are tremendous on cosmological scales, their influences in a laboratory on the earth are weak and difficult to detect. This is where the concept of analog models becomes relevant \cite{R39,Xu:21,PhysRevX.4.011038, R60}. Analog models have gained prominence in physics and mathematics, offering new perspectives and enabling the exchange of ideas between different  scientific disciplines \cite{doi:10.1080/09500340.2013.769638}. In recent years, there have been numerous attempts to employ physical systems, as analogy platforms to simulate and explore some aspects  of general relativity theory in laboratory \cite{Xu:18,R29,R48,PhysRevA.78.063814,Tavakoli2018}.
 For instance, analog models have been used to understand black hole physics, leading to the development of a thermodynamic theory for these objects and the discovery of the Hawking effect \cite{R16}.
Furthermore, several types of analog physical systems  on two-dimensional curved surfaces have been proposed to  model the general relativity. Given a constant time and extracting the equatorial slice of Friedmann-Robertson-walker’s   space-time, we arrive at a two-dimensional curved surface with constant curvature. In these types of analog models, a curved space is created by engineering the geometry of the space itself to investigate its effects on certain physical systems \cite{doi:10.1080/23746149.2020.1759451,R61,PhysRevD.109.065025}.
On the other hand, quantum thermodynamics  \cite{binder2018thermodynamics,R56} has attracted considerable attention for elucidating the fundamental relationship between quantum mechanics and thermodynamics \cite{PhysRevE.95.032111}.
The conversion of energy into mechanical work is essential for almost any industrial process \cite{R1}.
Quantum thermodynamics investigates fundamental concepts such as temperature, heat, and work in the context of quantum physics. This field primarily focuses on studying thermal machines that  operate in the quantum realm and determining how quantum properties can be exploited  to enhance their efficiency.
Quantum heat engines, which attracted much attention in recent year, are designed based on quantum thermodynamic cycles with different working substances, such as spin systems \cite{LIU201469, PhysRevE.104.054128}, single ions \cite{PhysRevE.95.032111}, harmonic-oscillators \cite{PhysRevE.79.041129,PhysRevE.91.062134,CHATTOPADHYAY2021126365},  quantum Brownian oscillators \cite{PhysRevE.88.012130}, two-level systems~\cite{PhysRevE.96.032110, PhysRevE.103.032130, PhysRevE.85.041148}, single-particles \cite{e18050168,PhysRevLett.123.240601,gelbwaser2018single},  two-particles \cite{PhysRevResearch.5.013088}, multilevel systems \cite{PhysRevE.72.056110,Uzdin_2014}, optomechanical systems \cite{PhysRevA.90.023819}, and cavity quantum electrodynamics systems \cite{PhysRevE.84.041127}. Among these, the so-called quantum  Otto engine  is presented, consisting of a quantum working substance manipulated between two heat reservoirs under two adiabatic and two isochoric branches \cite{R21,kieu2006quantum,e19040136}.

To address the question of the efficiency of a quantum heat engine,  compared to its classical counterpart, several protocols have been proposed and analyzed~\cite{R43}.
 In \cite{R42}, it is shown that non-thermal baths, such as coherent bath \cite{doi:10.1073/pnas.1110234108} and squeezed bath \cite{PhysRevLett.112.030602}, may impart not only heat, but also mechanical work to a  heat engine. While such engines can exceed the Carnot efficiency, they do not violate the second-law of thermodynamics\cite{Levy2018}.

 In Ref.~\cite{PhysRevD.109.065025}, by using a delta-coupled Unruh-Dewitt detector in Minkowski space-time, a relativistic quantum Otto engine is considered to extract work from a quantum field in a thermal state.  In Ref.~\cite{Kollas2024}, quantum Otto refrigeration cycles are constructed by utilizing the gravitational redshift experienced by photons in curved spacetime. The performance of a quantum Otto engine with a working medium of a relativistic particle is analyzed in Ref.~\cite{R20}. Additionally, the thermal properties of a quantum heat engine with a relativistic moving heat bath  are investigated in Ref.~\cite{R10}.
 In a related study, a quantum Otto engine was introduced, in which the effective temperatures observed by stationary moving observers are utilized \cite{R8}. Also, the amount of work extracted by an observer moving in a circular path is investigated in Ref.~\cite{R9}.
 Further research has investigated how relativistic energies and space-time geometry affect the thermal efficiency of thermodynamic cycles~\cite{R7}. However, the impact of curvature on the performance of quantum thermal engines remains a relatively unexplored topic.

In the present contribution, we adopt an analog model of general relativity to investigate the effects of spatial curvature  on a  quantum Otto engine.
We consider a quantum Otto cycle with a quantum harmonic oscillator on a circle as it’s  working substance. We assume that two classical hot and cold thermal baths are located at places with different curvatures. We calculate the curvature-dependent work and heat in our Otto  cycle and study the curvature effects on it's thermal efficiency. It is worth noting that the properties of the harmonic oscillator on a circle depend on the  curvature of the circle, and therefore, making it a suitable platform to investigate the spatial  curvature effects on this thermodynamic process. Moreover, this analog model offers the advantage of analytical tractability, facilitating a detailed exploration of curvature-induced phenomena.

The paper is organized as follows. In Section II, we provide a brief review of the quantum harmonic oscillator on a circle. In Section III, after a concise overview of the quantum Otto’s  engines, we propose our curvature-dependent scheme of a  quantum Otto engine with a  harmonic oscillator on a circle as the working substance.
Section IV is devoted to investigating  the efficiency of our engine, as well as its limiting case of small curvature (near the Earth surface) and large curvature (near massive celestial bodies like black holes).
In Section V, utilizing the theoretical framework established in the previous sections, we present our findings on the impact of curvature on the efficiency of  our  quantum Otto’s  engine.
 Finally the summary and concluding remarks are given in Section  VI.
\section{ Quantum Harmonic Oscillator on Circle} \label{sec2}
Quantum heat engines  generate work by utilizing quantum systems as their working substance. To investigate the spatial curvature effects on a quantum Otto’s  engine, in this paper, we assume that the working substance is  a quantum harmonic oscillator on a circle. Therefore, we begin with a brief introduction to the quantum harmonic oscillator on the circle.

\begin{figure}
\centering
  \includegraphics[width=0.5\textwidth]{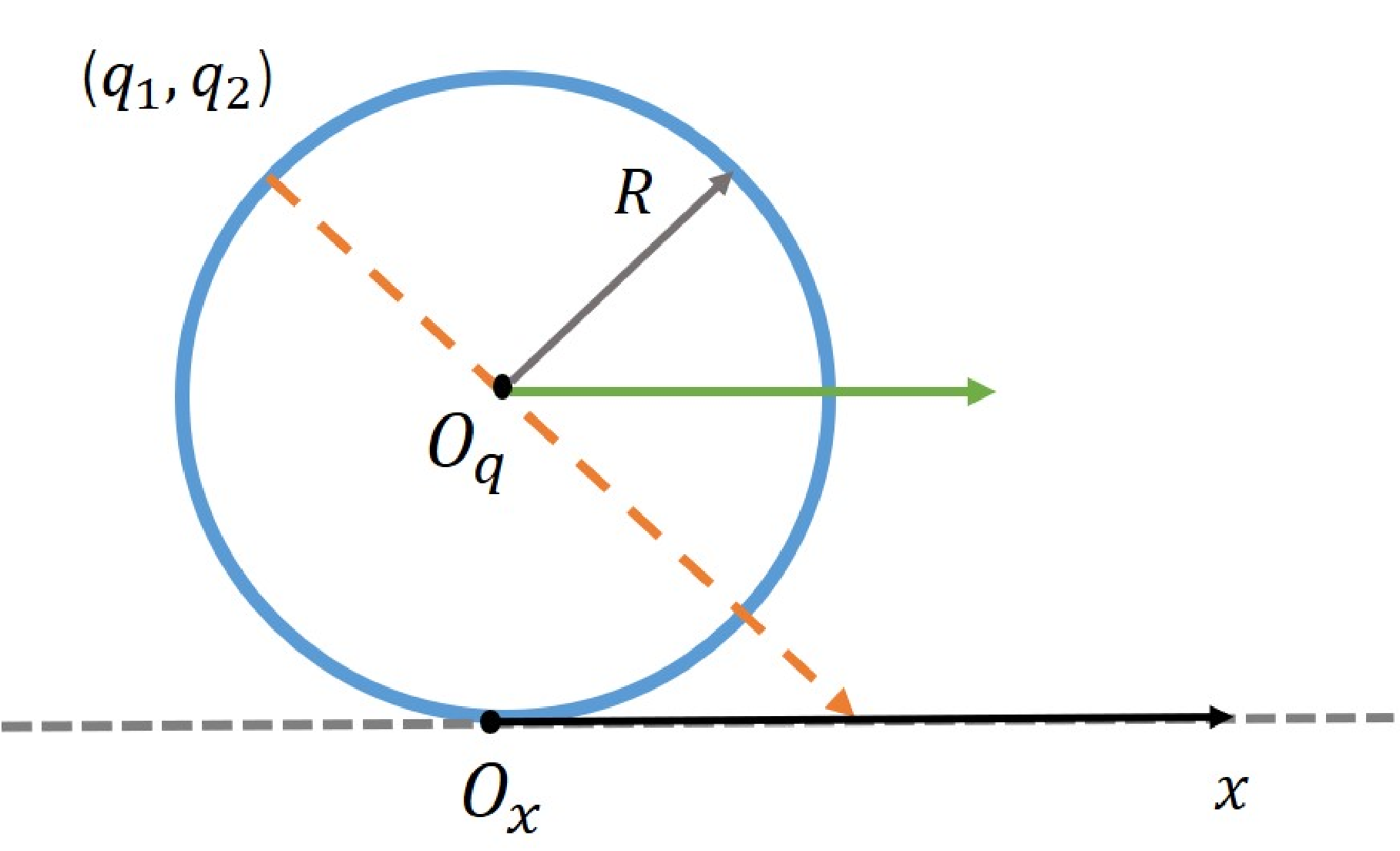}
\caption{ Coordinate systems and the gnomonic projection from a circle with radius $R$ onto a line.}
 \label{fig:1}
\end{figure}

Recently, some of the authors of this paper studied the quantum dynamics of a harmonic oscillator constrained on a circle with radius $R$ \cite{Mahdifar:2021owk}.
By employing the gnomonic projection, which is a
projection onto the tangent line from the center of the circle and denoting the Cartesian coordinate of this projection by $x$ (see Fig.~\ref{fig:1}), the Hamiltonian of the quantum harmonic oscillator on a circle is derived as follows:
 \begin{equation}\label{3}
 \hat{H} ({\lambda}) = \frac{1}{2} \Big[-(1+\lambda x^{2})^{2} \frac{d^{2}}{dx^{2}} - 2\lambda x (1+\lambda x^{2})\frac{d}{dx}\Big] +\frac{1}{2} x^{2},
\end{equation}
where $\lambda = 1/R^{2}$ is the curvature of the circle and the natural system of units is employed ( $ m= \hbar = \omega = 1$). The energy eigenvalues of the quantum oscillator on a circle are obtained as:
\begin{equation}\label{2}
E_{n}(\lambda) = \gamma (n + \frac{1}{2}) + \frac{\lambda}{2}  n^{2},
\end{equation}
where
\begin{equation}\label{formula_label}
 \gamma  = \frac{ (\lambda + \sqrt{{\lambda}^{2} + 4} )}{2}.
\end{equation}
As it is seen, the energy spectrum \eqref{2} is expressed as a function of the spatial curvature $\lambda$.
It is obvious that in the limit of $\lambda\rightarrow 0$, $\gamma$ approaches unity, and we recover the energy eigenvalues of the quantum harmonic oscillator on a straight line: $E_{n}(\lambda = 0) = (n + \frac{1}{2})$.
\begin{figure}
\centering
\includegraphics[width=0.5\textwidth]{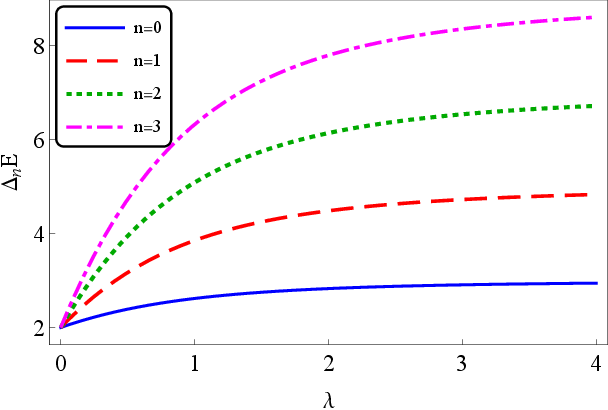}
\caption{ Dimensionless energy gaps as a function of the spatial curvature parameter $\lambda$, for $n = 0, 1, 2, 3$.}
\label{fig:2}
\end{figure}
It is worth noting that the spatial curvature parameter, which introduces a nonlinear character to the quantum oscillator on a circle, manifests in the anharmonicity of the energy spectrum. In other words, when we go from a flat line to a curved circle, the energy levels are no longer evenly spaced. Therefore, our quantum oscillator  exhibits an effective piston-like behavior through variations in the curvature parameter  $\lambda$.
In Fig. \ref{fig:2}, we have plotted the dimensionless variation of energy gaps:
  \begin{equation}\label{formula_label}
\triangle_{n}E(\lambda)  = \frac{E_{n+1}(\lambda)-E_{n}(\lambda)}{E_{0}(\lambda)},
\end{equation}
as a function of the spatial curvature parameter. It is seen that for a fixed value of $\lambda$ $(\lambda\neq 0)$, the energy gaps  increase with increasing the value of  $n$. Furthermore, with increasing the curvature $\lambda$, the uniformity of the energy gaps between the eigen-energies diminishes further.
\section{QUANTUM CURVATURE-DEPENDENT OTTO ENGINE }\label{sec3}
\begin{figure}[H]
\centering
  \includegraphics[width=0.6\textwidth]{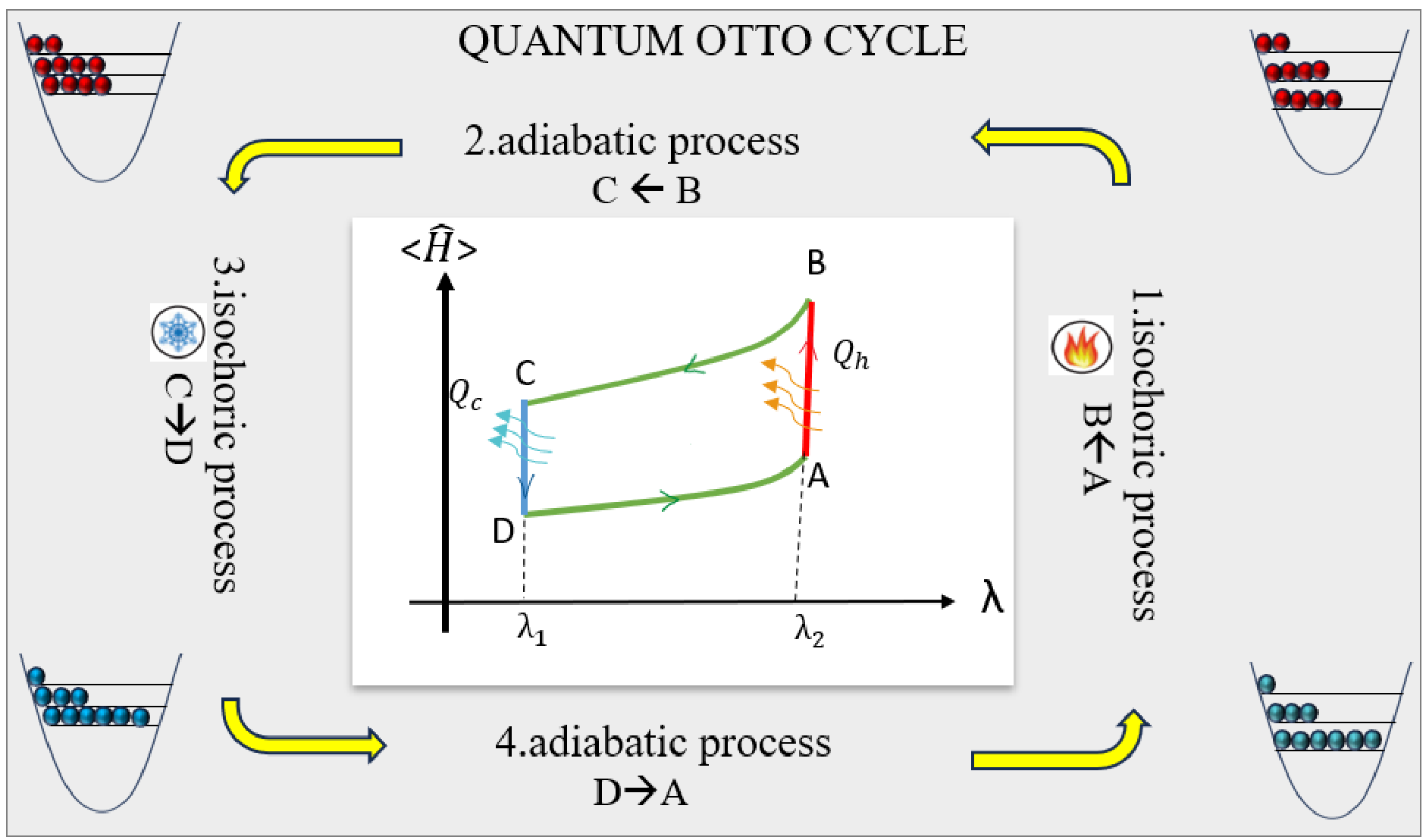}
\caption{ An illustration of the mean energy-the spatial curvature, diagram of a quantum Otto cycle based on
the harmonic oscillator on a circle }
 \label{fig:3}
\end{figure}

In this section, we consider a quantum  Otto’s  engine  with a harmonic oscillator on a circle as its working substance, as
 illustrated in Fig. \ref{fig:3}. This engine consists of two isochoric branches, one with a hot bath and the other with a cold thermal bath  at two different temperatures   $T_{h}>T_{c}$ \cite{PhysRevE.93.022122, PhysRevE.100.032144}. A quantum isochoric process (constant volume process) is similar to its classical isochoric process. In contrast, a classical adiabatic thermodynamic process does not always require the occupation probabilities to remain constant, while a quantum adiabatic process proceeds slowly enough so that the generic quantum adiabatic condition is satisfied, and keeps the population distributions constant \cite{PhysRevE.76.031105}. Moreover, during quantum adiabatic stages, the working substance Hamiltonian  varies while population distribution remains constant \cite{PhysRevE.108.054103}. It is worth mentioning that if this variation in the Hamiltonian causes the energy gaps to scale at a constant ratio,
the efficiency of the engine is like its classical counterpart.
However, if the deformation of the Hamiltonian leads to
inhomogeneous energy gaps, the heat engine’s efficiency
can exceed its classical counterpart \cite{PhysRevLett.120.170601}.
The first step in understanding the  thermodynamics of a harmonic oscillator on a circle is to find  the correct expression for the equilibrium distribution when the system is  in contact with a thermal bath. The equilibrium distribution of a quantum oscillator on a circle, weakly coupled to a large classical bath, will eventually reach the corresponding Gibbs thermal equilibrium state, $ \rho =  e ^{- \hat{H}(\lambda)/T}/Z(\lambda)$, \cite{R59}
where the canonical partition function of the quantum oscillator, acting as the working substance of our Otto’s  engine, is given by
\begin{equation}\label{formula_label}
\ Z(\lambda) =  \sum_{n}  e^{-\frac{E_{n}(\lambda)}{T}}.
\end{equation}
Here, we take $k_{B}$ as unity and $E_{n}(\lambda)$ is given by \eqref{2} .

It should be note that in our analog model of the quantum Otto’s  cycle, we assume that the hot and cold thermal bath are located at two different heights above  the ground. In other words, to investigate the curvature effects of the physical space on the work done by the cycle, and the heat engine's efficiency, we  place the hot thermal bath in a location with greater spatial curvature than the cold thermal bath, i.e. $\lambda_{2}\geq \lambda_{1} $.
However, we study these curvature effects  using an analog model, presented by a quantum harmonic oscillator on a circle.

 The internal energy is simply determined by the expectation value of the Hamiltonian over the state of the working substance as:
\begin{equation}\label{formula_label}
 \langle \hat{H}(\lambda) \rangle= \sum_{n}P_{n}(\lambda)E_{n}(\lambda),
\end{equation}
where the occupation probability  of the system, $ P_{n}(\lambda)$, in thermal equilibrium with the bath  is given by
\begin{equation}\label{formula_label}
\ P_{n} (\lambda)= \frac{e^{-\frac{E_{n}(\lambda)}{T}}}{Z(\lambda)}.
\end{equation}
\subsection{Processes of a quantum Otto engine cycle }
The quantum Otto cycle consists of four distinct stages \cite{PhysRevA.105.022609}.
In order to evaluate the performance of our quantum Otto’s  engine, we need to calculate the work and the heat during each stage. Heat is exchanged with the baths
during the isochoric thermalization in the first and third stages, while work is performed during two and four unitary strokes \cite{PhysRevE.103.032130}.

\subsubsection{Hot isochore $(A\rightarrow B)$}

 During step 1, the quantum working substance is connected to the heat bath at temperature  $T_{h}$, converting  from state $ A $ to  $ B $, with equal energy levels $ E_{n}^{B}(\lambda_ {2})= E_{n}^{A}(\lambda_ {2}) $. Moreover, the energy level gaps remain unchanged as the system absorbs energy, and the temperature of the working substance raises until it reaches thermal equilibrium with the hot bath.  At the end of this isochoric process, the occupation probabilities in each eigenstate satisfies the Boltzmann distribution. Thus, the heat  absorbed by the working substance from the hot bath is given by:
\begin{equation}\label{1}
Q_{h}=\sum_{n}E_{n}(\lambda) dP_{n}(\lambda)= \sum_{n} E_{n}^{A}(\lambda_{2}) \Big[P_{n}^{B}(\lambda_{2})- P_{n}^{A}(\lambda_{2})\Big].
\end{equation}
Furthermore, since the curvature is equal to $\lambda_{2}$, no work is done during this step, and we have:
$$W_{A \rightarrow B}=0.$$

\subsubsection{Isentropic  expansion  $(B\rightarrow C)$}


After decoupling from the bath, the system undergoes a quantum adiabatic transformation from state  $B$, with curvature $\lambda_{2}$, to state $C$, with  curvature $\lambda_{1}$. In quantum adiabatic processes, the adiabatic condition  is met when changes occur slowly enough to keep population distributions constant $ (dP_{n}=0) $ and no heat is exchanged $({d\hspace*{-0.08em}\bar{}\hspace*{0.1em}} Q = 0)$. However, work can still be done during this process despite the lack of heat exchange. This transformation ensures that the probabilities in state  $C$ match those in state $ B $, i.e.$,  P_{n}^{C}(\lambda_ {1})=  P_{n}^{B}(\lambda_ {2}) $, and the energy eigenvalues are adiabatically transferred from  $ E_{n}^{B}(\lambda_ {2})= E_{n}^{A}(\lambda_ {2})$ to $ E_{n}^{C}(\lambda_ {1})$. Therefore, we have:
$$\\Q_{B \rightarrow C}=0, $$
and
\begin{equation}\label{formula_label}
W_{B \rightarrow C} =\sum_{n} P_{n}dE_{n} =\sum_{n} P_{n}^{B}(\lambda_{2}) \Big[ E_{n}^{A}(\lambda_ {2}) - E_{n}^{C}(\lambda_{1})\Big].
\end{equation}

\subsubsection{ Cold isochore $(C\rightarrow D)$}

In the third step, the system  cools down to  point $D$, where the eigen-energy of the working substance remains the same as  point $ C $, i.e.$, E_{n}^{D}(\lambda_{1}) =E_{n}^{C}(\lambda_{1}) $, while the population adjusts to  $P_{n}^{D}(\lambda_{1})$.
 The heat exchange between the substance and the cold bath, as well as the work done during the isochoric cooling can be calculated as:
\begin{equation}\label{formula_label}
Q_{c}=\sum_{n} E_{n}^{C}(\lambda_{1}) \Big[P_{n}^{D}(\lambda_{1})- P_{n}^{C}(\lambda_{1})\Big],
\end{equation}
$$ W_{C\rightarrow D}=0.$$
\subsubsection{Isentropic compression $(D\rightarrow A)$}

Ultimately, the system  resets to its initial point $A$, by completing the cycle with another quantum adiabatic transformation, provided that $ P_{n}^{A}(\lambda_{2}) = P_{n}^{D}(\lambda_{1}) $. In this adiabatic step, there is no heat exchange between the system and its surroundings, and the work can be directly determined by:
\begin{equation}\label{formula_label}
W_{D \rightarrow A} =\sum_{n} P_{n}^{D}(\lambda_{2}) \Big[E_{n}^{D}(\lambda_ {2}) - E_{n}^{A}(\lambda_{1})\Big].
\end{equation}
Also we have:
$$\\Q_{D \rightarrow A}=0. $$
\subsection{Work and heat of the quantum Otto cycle }
Now we can obtain the net work performed during the cycle \cite{R59} as follows:
 \begin{eqnarray}\label{12}
 W & = & -(W_{B\rightarrow C} + W_{D\rightarrow A}) = Q_{h} -Q_{c} \\
      & = & \sum_{n} \Big[E_{n}^{A}(\lambda_{2}) - E_{n}^{C}(\lambda_{1})\Big] \Big[P_{n}^{B}(\lambda_ {2}) - P_{n}^{A}(\lambda_{2})\Big]. \nonumber
\end{eqnarray}
It is worth noting that the factor $ E_{n}^{A}(\lambda_{2}) - E_{n}^{C}(\lambda_{1}) $ represents the variation of the energy eigenvalue during quantum adiabatic transformations. We can evaluate work using the thermal population distribution    $P_{n}^{B}(\lambda_ {2})= P_{n}^{C}(\lambda_{1}) $   and  $P_{n}^{D}(\lambda_ {1})= P_{n}^{A}(\lambda_{2})$. From this analysis, it should be evident that the quantum Otto’s  cycle is an irreversible process.
\section{ THE THERMAL EFFICIENCY OF QUANTUM OTTO ENGINE }\label{sec5}

The first law of thermodynamics explains the conversion of heat, a form of energy, into work, while the second law determines how much of this heat can be effectively converted into work. Efficiency is defined as the ratio of output energy  to input energy  \cite{Abah_2017}. The  thermal efficiency of our Otto cycle is defined by
\begin{equation}\label{13}
\eta= \frac{W}{Q_{h}},
\end{equation}
where the heat injected into the system from the hot bath, $Q_{h}$,  is  given by equation \eqref{1}.
\subsection{ LIMITING CASE OF THERMAL EFFICIENCY}
In this section, we consider two limiting cases of our curvature-dependent Otto engine’s
thermal efficiency \eqref{13}: the small spatial curvature limit and the large spatial curvature limit.
In these limiting cases, we assume that the curvature difference between  the position of   the cold and the hot bath,  denoted by $\varepsilon$  is too small, so that  $\lambda_{1}= \lambda- \varepsilon $ and  $\lambda_{2}= \lambda$.
\subsubsection{The small spatial curvature limit}

Here, we calculate the efficiency in the small spatial curvature limit, such as, near the Earth, assuming a slight difference in spatial curvatures and a small temperature difference between the two heat sources.
 Defining the temperature difference between two baths as $( \theta = T_{h} - T_{c} )$, and expressing the temperature of the hot bath in terms of the cold bath  by $\frac{1}{T_{h}} = \frac{1}{T_{c}}-\frac{\theta}{T_{c}^{2}}$,
we can then write
\begin{equation}\label{formula_label}
W \simeq \frac{\varepsilon \theta }{Z_{A}(\lambda) T_{A}^{2}} \sum_{n=0}^{\infty}  E'_{n}(\lambda) E_{n}(\lambda) e^{-\frac{E_{n}(\lambda)}{T_{A}}},
\end{equation}
and
\begin{equation}\label{formula_label}
Q_{h} \simeq \frac{\theta}{Z_{A}(\lambda) T_{A}^{2}} \sum_{n=0}^{\infty}  E_{n}^{2}(\lambda) e^{-\frac{E_{n}(\lambda)}{T_{A}}}.
\end{equation}
Here, we assume that, in this limiting case, $Z_{A}(\lambda) \simeq Z_{B}(\lambda)$, and
$E'_{n}(\lambda)$ refers to the derivative of Eq.~\eqref{2} with respect to $\lambda$. For small value of $\lambda$, this derivative is approximately equal to $(n^{2}+n+\frac{1}{2})/2$.
Now, by using these approximate relations, the following compact form is obtained for our engine's efficiency:
\begin{equation}\label{formula_label}
\eta_{s}\simeq\frac{\varepsilon E'_{0}}{E_{0}}(1+\frac{E_{1}E'_{1}}{ E_{0} E'_{0}} e^{-\frac{E_{1}-E_{0}}{T_{A}}}).
\end{equation}
 For small curvature differences, such as  $\lambda_{2}= 0.011$,  $\lambda_{1}= 0.01$ and $\theta=0.05$, the engine’s  approximate efficiency  $\eta_{s} \simeq 0.001488$  closely matches the exact value from Eq.~\eqref{13}:  $\eta=0.001314$. It turns out that the two values are in near agreement.
\subsubsection{The large spatial curvature limit}

To investigate thermal efficiency when the spatial curvature takes on its most extreme possible values, such as near a super massive black holes, we assume in our analog model that two values of $\lambda_{1}$ and  $\lambda_{2}$, are nearly large, but their difference remains small. Therefore, the change in eigenstate energies can be written as:
\begin{equation}\label{formula_label}
\Delta E_{n}= E_{n}^{A}(\lambda)- E_{n}^{C}(\lambda - \varepsilon) \cong (n+1)^{2}.
\end{equation}
 Considering that  $\lambda$ approaches large values, it can be easily shown  that $\gamma \simeq \lambda $. To carry out our study, we need the Jacobi theta function to calculate the expansion coefficients,  which is defined as \cite{weisstein2000jacobi}:
\begin{eqnarray}\label{20}
\vartheta_{3}(z,q)=1+ 2 \sum_{n=1}^{\infty} q^{n^{2}} \cos(2 n z).
\end{eqnarray}
By choosing  $z=0$ and $q=e^{-\frac{\lambda}{2 T}}$, the following relations hold:
\begin{equation}\label{21}
\sum_{n=0}^{\infty} q ^{(n+1)^2}= \frac{\vartheta_{3}(q)-1}{2},
\end{equation}
\begin{equation}\label{22}
\sum_{n=0}^{\infty} (n+1)^{2} q ^{(n+1)^2}= \frac{q}{2} \vartheta'_{3}(q),
\end{equation}
where the prime indicates a derivative with respect to its argument.
Using Eq. \eqref{21}  and \eqref{22}, the harvested work and the heat absorbed during our quantum heat engine cycle are respectively,  written as
\begin{eqnarray}\label{formula_label}
W=\frac{\varepsilon}{2}[\frac{q_{h}\theta'_{3}(q_{h})}{\theta_{3}(q_{h})-1}-\frac{q_{c}\theta'_{3}(q_{c})}{\theta_{3}(q_{c})-1}], \nonumber
\end{eqnarray}
\begin{eqnarray}\label{formula_label}
Q_{h}= \frac{\lambda}{2}\Big[\frac{q_{h}\theta'_{3}(q_{h})}{\theta_{3}(q_{h})-1}-\frac{q_{c}\theta'_{3}(q_{c})}{\theta_{3}(q_{c})-1}\Big], \nonumber
\end{eqnarray}
where $q_{h}=e^{-\frac{\lambda}{2 T_{h}}}$ and $q_{c}=e^{-\frac{\lambda}{2 T_{c}}}$.
Bearing these in mind, it is easily seen that the efficiency of our  quantum heat engine,  up to the leading order of $\varepsilon$, is given by:
\begin{eqnarray}\label{formula_label}
\eta_{l} \simeq \frac{\varepsilon}{\lambda}.
\end{eqnarray}
For example, if we set $\lambda_{1}=9.8$  and $ \lambda_{2}=10$, the approximate efficiency of the  engine, i.e., $\eta_{l}=\frac{\varepsilon}{\lambda}=0.02$, is very  close to the exact value of $\eta= 0.0197$.
 In this  limiting case, we see that the engine's efficiency is independent of temperature.
 \section{RESULTS AND REMARKS}\label{sec4}
In this section, we proceed to study the effects of curvature  on our quantum Otto’s engine  using the analog model of a simple quantum harmonic oscillator on a circle. As shown in Eq. \eqref{12}, the work extracted and thus the heat engine’s efficiency of our proposed quantum Otto’s heat engine  depend on the net variation of the energy eigenvalues between points $A$ and $C$, as well as the population difference between points $A$ and $B$.
\begin{figure}
\centering
\includegraphics[width=0.5\textwidth]{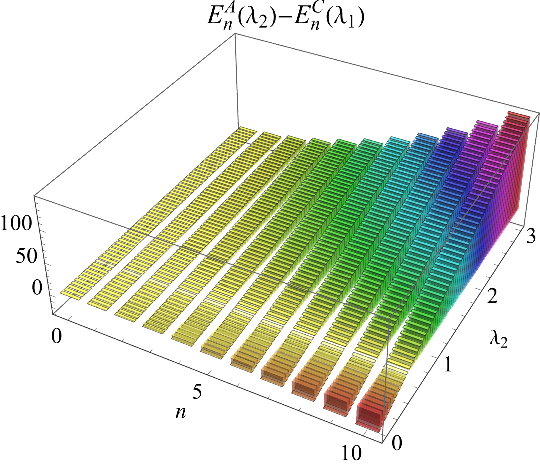}
\caption{  $E_{n}^{A}(\lambda_{2})-E_{n}^{C}(\lambda_{1})$ during
the adiabatic transformations versus $n$ and $\lambda_{2}$ for $T_{h} = 1$ and $T_{c} = 0.1 $ .}
 \label{fig:4}
\end{figure}

More precisely, during the quantum adiabatic process the stage $2$ of the quantum Otto’s cycle, the curvature of space changes from $\lambda_{2}$ to $\lambda_{1}$, the structure of the energy levels varies from $E_{n}^{A}(\lambda_{2})$
to $E_{n}^{C}(\lambda_{1})$ and as a result, the distance between the energy levels
gaps, $E_{n}^{A}(\lambda_{2})-E_{n}^{C}(\lambda_{1})$, changes. In contrast, during the quantum isentropic process in the stage $1$,  the interaction of the system with the heat source changes, the population differences, $P_{n}^{B}(\lambda_ {2}) - P_{n}^{A}(\lambda_{2})$. As is evident, these changes depend on the curvature of space.

In Fig. \ref{fig:4},  we have plotted the variation of energy eigenvalue, $E_{n}^{A}(\lambda_{2})-E_{n}^{C}(\lambda_{1})$,
versus $\lambda_{2}$ and $n$ for  $\lambda_{1}=0.5$.
 It is seen that the variation of energy
eigenvalues between the point $A$ (with the curvature $\lambda_{2}$) and the point $C$
(with the curvature $\lambda_{1}$) increases with enhancement of $n$. Moreover, $E_{n}^{A}(\lambda_{2})-E_{n}^{C}(\lambda_{1})$ is
negative when $\lambda_{2}<\lambda_{1} $,  zero when $\lambda_{2}=\lambda_{1} $, and
positive when $\lambda_{2}>\lambda_{1}$.
\begin{figure}
\centering
\includegraphics[width=0.5\textwidth]{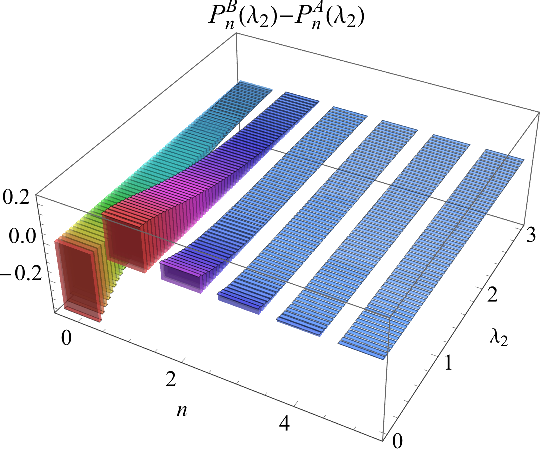}
\caption{ $P_{n}^{B}(\lambda_ {2}) - P_{n}^{A}(\lambda_{2})$ versus $n$ and $\lambda_{2}$  for $T_{h} = 1$ and $T_{c} = 0.1 $ .}
 \label{fig:5}
\end{figure}
In Fig. \ref{fig:5}, we display the population differences, $P_{n}^{B}(\lambda_ {2}) - P_{n}^{A}(\lambda_{2})$, as  functions of $\lambda_{2}$ and $n$ for the hot
bath temperature $T_{h} = 1$ and the cold bath temperature $T_{c} = 0.1$. As is seen, the population differences decrease with increasing the value
of $n$ and also increasing the curvature of space $\lambda_{2}$. This indicates that from $A$ to $B$, the population differences for a fixed value of $n$ are
greater in flat space than in curved space.
\begin{figure}
\centering
  \includegraphics[width=0.6\textwidth]{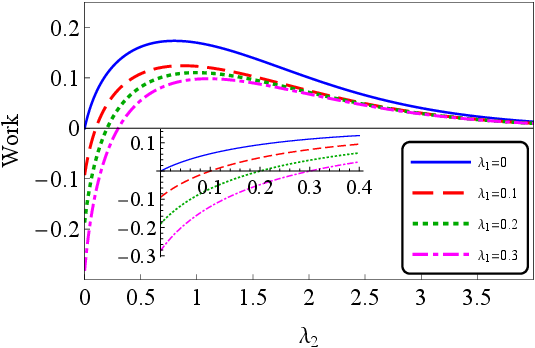}
\caption{Extracted work versuse the spatial curvature of the
working substance $\lambda_{2}$ for  $(T_{h} = 1)$ and  $(T_{c} = 0.1)$. }
 \label{fig:6}
\end{figure}
Fig. \ref{fig:6} displays the total  work extracted from the quantum Otto engine as a function of the spatial curvature  $\lambda_{2}$ in which the hot bath is located, for several values of the curvature of  the cold bath location, with $T_{h} =1$ and  $T_{c} = 0.1$.

 A fundamental condition for the performance of a heat engine is the positive  work extraction: $W_{ext}\geq 0$. The inset of Fig. \ref{fig:6} clearly illustrates that, the positive work can be extracted from our quantum Otto engine, provided that: $\lambda_{2} > \lambda_{1}$ . When there is no change in the curvature of space, i.e., $\lambda_{2}=\lambda_{1}$, there is also no change in the energy eigenvalues between $A$ and $C$ and the work extracted   from the quantum Otto machine is zero.
Also, when $\lambda_{2}<\lambda_{1}$, the extracted work becomes negative.
 This shows the motivation behind our initial assumption of placing  the
hot reservoir in a more curved space than the cold reservoir.

It should be noted that by choosing $\lambda_{2} >\lambda_{1}$, the total extracted work first increases with the increase of $\lambda_{2}$ and then tends to zero. In addition, it is seen that the extracted work  decreases with the  increase of the curvature $\lambda_{1}$. Also, the peaks shift toward larger value of
$\lambda_{2}$ as  $\lambda_{1}$ increases.
\begin{figure}
\centering
  \includegraphics[width=0.6\textwidth]{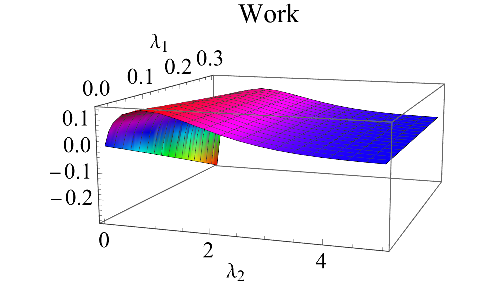}
\caption{The extracted work versus the spatial curvatures $\lambda_{1}$ and $\lambda_{2}$ for $T_{h} = 1$ and $T_{c} = 0.1 $.}
 \label{fig:7}
\end{figure}
In order to illustrate obtained results more clearly, in Fig. \ref{fig:7}  we have also displayed a three-dimensional plot of the extracted work versus the curvatures $\lambda_{1}$ and $\lambda_{2}$.
\begin{figure}
\centering
  \includegraphics[width=0.6\textwidth]{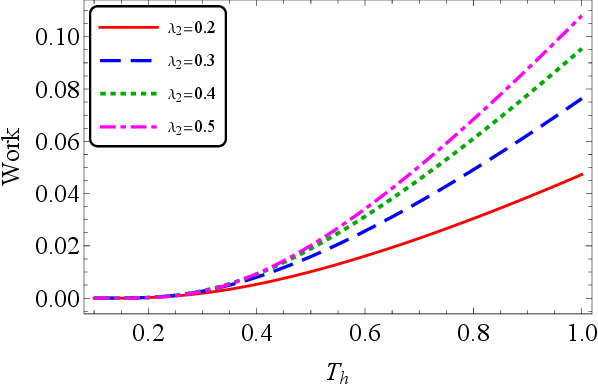}
\caption{The extracted work versus hot bath temperatures, $T_{h}$, for $T_{c} =0.1$ and $\lambda_{1}=0.1$.}
 \label{fig:8}
\end{figure}
In Fig. \ref{fig:8}, the extracted work is plotted  versus $T_{h}$ for different values of $\lambda_{2}$ for $T_{c}=0.1$ and $\lambda_{1}=0.1$.
The results clearly show that for a fixed value of  $\lambda_{2}$, the  extracted work is increased with increasing  the temperature difference between the hot and the cold bath. On the other hand, for a given temperature  $T_{h}$, increasing  the curvature $\lambda_{2}$  leads to an increase in the extracted work. In other words,  the increase in the difference between the curvatures   $\lambda_{1}$ and $\lambda_{2}$
(the height of the hot  bath and cold bath) plays a compensation role against the decrease in the difference between the  temperature  $T_{h}$ and  $T_{c}$.
\begin{figure}
\centering
  \includegraphics[width=0.6\textwidth]{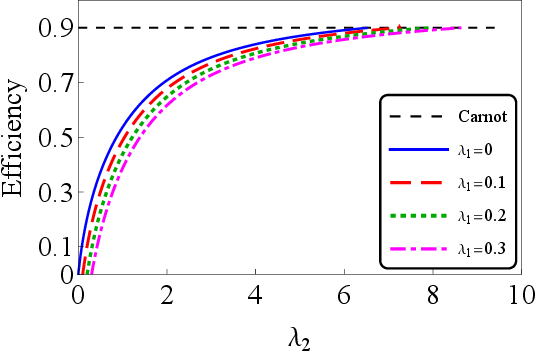}
\caption{The  quantum Otto engine’s
efficiency as a function of  $\lambda_{1}$ and  $\lambda_{2}$ for hot and cold bath temperatures $T_{h} =1$ and $T_{c} = 0.1$, respectively.}
 \label{fig:9}
\end{figure}

In Fig. \ref{fig:9}, we have plotted the heat engine’s efficiency  versus the spatial curvature, $\lambda_{2}$ for $T_{h} =1$, $T_{c} = 0.1$ and different values of $\lambda_{1}$.
This figure clearly show that for any fixed value of  $\lambda_{1}$, increasing spatial curvature of the hot bath location $\lambda_{2}$, leads to a rapid increase in the heat engine’s efficiency until it reaches a stable value.

One of the important factors in enhancing the efficiency of a quantum heat engine over it’s classical counterpart is that the change in the Hamiltonian of it’s working substance should be such that the difference between the energy gaps during the engine's irreversible process changes in an anharmonicity manner \cite{PhysRevLett.120.170601}.

It is worth noting that the  results of Fig. \ref{fig:9} indicate that the efficiency of our analog heat engine model depends on the spatial curvature of the working substance's location.
Therefore, by utilizing a quantum harmonic oscillator on a circle as the working substance, we can control  the work and efficiency of the Otto engine. Moreover, it is possible to enhance it’s efficiency  to the Carnot bound  $(\eta_{Carnot}=1-{T_{cold}}/{T_{hot}})$, by adjusting the curvature of space parameters $\lambda_{1}$ and  $\lambda_{2}$ (the height of the hot and cold baths).
 It is worth mentioning that despite the increase in efficiency, our quantum Otto engine remains bound by the Carnot limit and fully adheres to the second law of thermodynamics~\cite{Levy2018}.

On the other hand, for any fixed value of  $\lambda_{2}$, increasing the value of $\lambda_{1}$ and thus, decreasing the distance between  hot and cold baths, leads to a decrease in heat engine’s efficiency. A similar result is also observed in a quantum Otto heat engine with a relativistically moving thermal bath. When the hot bath  moves at a relativistic speed relative to the cold bath, the distance between the two sources contracts, and thus, the heat engine is less efficient \cite{R34}.

\begin{figure}
\centering
  \includegraphics[width=0.6\textwidth]{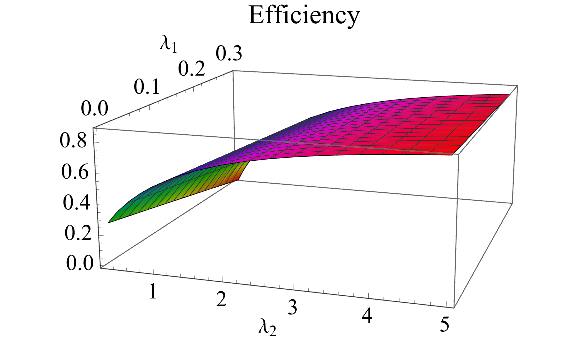}
\caption{The efficiency of quantum Otto cycle versus  $\lambda_{1}$ and  $\lambda_{2}$ for $T_{h} =1$ and $T_{c} = 0.1$ }
 \label{fig:10}
\end{figure}

Finally, to better visualize the results that have been obtained,  we display in Fig. \ref{fig:10} the variation of our engine efficiency  versus the curvatures $\lambda_{1}$ and $\lambda_{2}$.
\begin{figure}
\centering
  \includegraphics[width=0.6\textwidth]{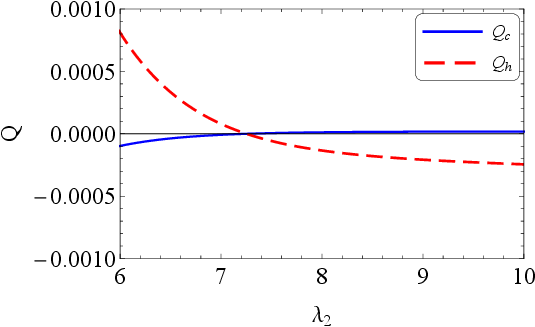}
\caption{ The heats $Q_{h}$ and  $Q_{c}$ interchanged
between the system and the thermal baths versus $\lambda_{2}$ for $\lambda_{1}=0.1$. }
 \label{fig:11}
\end{figure}

\emph{Remark}:
In our curvature-dependent quantum Otto engine, once the Carnot efficiency is achieved, further increasing the curvature difference between the two thermal baths leads to a transition from heat engine operation to refrigeration. For instance, Fig. \ref{fig:11} illustrates the heat exchanged between the working substance and both the cold and hot baths as a function of $\lambda_{2}$, with $\lambda_{1}=0.1$. As shown, when $\lambda_{2}$ exceeds approximately $7.2$, the heat exchanged with the hot bath becomes negative, while the heat exchanged with the cold bath remains positive, signifying the transition from a heat engine to a refrigerator.

 \section{SUMMARY AND CONCLUDING REMARKS}\label{sec6}
 In this paper,  we have considered a quantum Otto’s engine with a quantum  harmonic oscillator on a circle as it’s working substance, serving as an analog model of general relativity. By utilizing the curvature-dependent energies of this oscillator, we have investigated how the curvature effects of the physical space affects the properties of our quantum heat engine. Assuming that two thermal baths are located at places with different curvatures, we have calculated the curvature-dependent work and heat, with a particular focus on the effects of curvature on it’s thermal efficiency. The result show that by adjusting the curvature difference between the bath's locations, we can control  the efficiency of our heat engine, potentially allowing it to reach the Carnot efficiency limit.

\bmhead{Acknowledgements}

We gratefully acknowledge support for this work by the Office of Graduate Studies and Research of Isfahan for their support. We would also like to extend our thanks to Shahrekord University for  their assistance.


\end{document}